\documentclass[sigconf]{acmart}
\usepackage{pifont}
\usepackage{multirow}
\usepackage{enumitem}
\usepackage{subfigure}
\usepackage{threeparttable}
\usepackage{listings}
\usepackage{balance}
\usepackage{hyperref}
\usepackage{url}
\usepackage[hyphenbreaks]{breakurl}
\hypersetup{breaklinks=true}

\newcommand{\M}{In-UCDS}

\newcommand{\UOFMatric}{$\mathcal{M}_{UOF}$}

\AtBeginDocument{%
  \providecommand\BibTeX{{%
    \normalfont B\kern-0.5em{\scshape i\kern-0.25em b}\kern-0.8em\TeX}}}

\begin{document}

\title{Reproducibility Companion Paper:\\In-processing User Constrained Dominant Sets for User-Oriented Fairness in Recommender Systems}

\author{Yixiu Liu}
\affiliation{%
  \institution{School of Cyberspace, Hangzhou Dianzi University}
  \city{Hangzhou}
  \country{China}
}
\email{liuyixiu@hdu.edu.cn}

\author{Zehui He}
\affiliation{
    \institution{School of Cyberspace, Hangzhou Dianzi University}
  \city{Hangzhou}
  \country{China}
}
\email{23270225@hdu.edu.cn}

\author{Yuyuan Li}
\orcid{0000-0003-4896-2885}
\authornote{Corresponding author.}
\affiliation{%
  \institution{School of Communication Engineering, \\Hangzhou Dianzi University}
  \city{Hangzhou}
  \country{China}
}
\email{y2li@hdu.edu.cn}

\author{Zhongxuan Han}
\orcid{0000-0001-9957-7325}
\affiliation{%
  \institution{College of Computer Science, Zhejiang University}
  \city{Hangzhou}
  \country{China}
}
\email{zxhan@zju.edu.cn}

\author{Chaochao Chen}
\orcid{0000-0003-1419-964X}
\affiliation{%
  \institution{College of Computer Science, Zhejiang University}
  \city{Hangzhou}
  \country{China}
}
\email{zjuccc@zju.edu.cn}

\author{Xiaolin Zheng}
\orcid{0000-0001-5483-0366}
\affiliation{%
  \institution{College of Computer Science, Zhejiang University}
  \city{Hangzhou}
  \country{China}
}
\email{xlzheng@zju.edu.cn}
\renewcommand{\shortauthors}{Yuyuan Li et al.}

\begin{abstract}
In this paper, we reproduce experimental results presented in our earlier work titled ``In-processing User Constrained Dominant Sets for User-Oriented Fairness in Recommender Systems'' that was presented in the proceeding of the 31st ACM International Conference on Multimedia.
This work aims to verify  the effectiveness of our previously proposed method and provide guidance for reproducibility. 
We present detailed descriptions of our preprocessed datasets, the structure of our source
code, configuration file settings, experimental environment, and the reproduced experimental results.
%
\end{abstract}




\maketitle

\section{Contribution Summary}\label{sec:summary}
Ensuring fairness in recommendation systems remains a challenge, especially for users with different activity levels, known as the user-oriented fairness problem.
To address this issue, we propose a novel fairness-aware training framework named In-processing User Constrained Dominant Sets (In-UCDS) that enhances disadvantaged users’ learning by leveraging constrained clustering and fairness-aware optimization.
Specifically, In-UCDS consists of two key stages: (1) UCDS modeling, where disadvantaged users are clustered with similar advantaged users via constrained dominant sets, and (2) in-processing training, where a fairness loss is formulated based on embedding differences and integrated into the backbone model to enhance fair learning.
Through these two stages, In-UCDS enables fairness-aware training without compromising recommendation performance.
Extensive experiments demonstrate the effectiveness of our approach. 

\section{ARTIFACTS DESCRIPTION}\label{sec:description}
The artifacts consist of datasets, source code, configuration files, and experimental environment settings.
%
%
\subsection{Dataset Preparation}\label{sec:dataset}
The datasets used in our experiments follow~\cite{Rahmani2022fairness}. Its source code and data are available at the following link\footnote{\url{https://github.com/rahmanidashti/FairRecSys}}.
Specifically, we use three real-world datasets from different domains, namely Epinion~\cite{massa2007trust}, MovieLens~\cite{harper2015movie}, and Gowalla~\cite{liu2017point}, to comprehensively evaluate the performance of the In-UCDS framework.
%
We adopt the widely used leave-one-out strategy for dataset splitting.
Specifically, each user’s validation and test set contains one randomly selected positive item and 99 negative items.
Additionally, we rank users based on their interaction frequency within each dataset and extract the top 5\% as advantaged users, while the remaining users are considered disadvantaged.
Our complete source code, including the datasets, is available on Baidu Netdisk and Google Drive:

\vspace{2.5pt}
\noindent{\texttt{\url{https://pan.baidu.com/s/1zNkoOw2R2PoFcepRvMIVzw?pwd=ppmb}}}

\vspace{2.5pt}
\noindent{\texttt{\url{https://drive.google.com/drive/folders/1L1tTwiRsuXU_pkDguPJA6BD_5IzJB0Fq?usp=sharing}}}

\vspace{2.5pt}
In the above links, the ``sigDatasets'' folder in the root directory contains the three datasets used in our experiments.
Each dataset follows the same format. Taking Epinion as an example, we provide the following details:
\begin{itemize}[leftmargin=*] \setlength{\itemsep}{-\itemsep}
    \item \textbf{groups}: contains two pre-split subfolders, "users" and "items". The "users" folder includes "active\_ids.txt" and "inactive\_ids.txt", while the "items" folder contains "longtail\_items.txt" and "shorthead\_items.txt".
    \item \textbf{Epinion\_data.txt}: the processed dataset used in our experiments.
    \item \textbf{Epinion\_train.txt}: the training dataset.
    \item \textbf{Epinion\_tune.txt}: the validation dataset.
    \item \textbf{Epinion\_test.txt}: the testing dataset.
    \item \textbf{ratings\_data.txt}: the raw dataset before preprocessing.
\end{itemize}

\subsection{Source Code Structure}
Our source codes in the main directory of "In-UCDS" include 3 folders and 6 files, which are available at: \url{https://github.com/hahhacx/In-UCDS}.
\begin{itemize}[leftmargin=*] 
    \setlength{\itemsep}{-\itemsep}
    \item \textbf{Folders:}
    \begin{itemize}
        \item \textbf{models/}: includes source code for recommendation models.
            \item \textbf{logs/}: stores the best performing model weights during training.
            \item \textbf{result/}: contains the test results of all the best performing models, saved in .csv format. Each row consists of four columns: user ids, item ids, predicted scores and labels.
    \end{itemize}

    \item \textbf{Files:}
    \begin{itemize}
        \item \textbf{main.py}: the main program for running the fairness model.
            \item \textbf{config.py}: stores configuration settings for the model training process.
            \item \textbf{myloss.py}: computes the loss values.
            \item \textbf{sigdatasets.py}: preprocesses the dataset.
            \item \textbf{test.py}: evaluates model performance metrics.
            \item \textbf{ucds.py}: implements the proposed method.
    \end{itemize}
\end{itemize}

In ``config.py'', we define dedicated configurations for four recommendation models: NeuMF~\cite{he2017neural}, PMF~\cite{sala2007prob}, VAECF~\cite{liang2018var}, and NGCF~\cite{wang2019neural}, collectively named ``<model>\_config''. Taking NeuMF~\cite{he2017neural} as an example, ``neumf\_config'' is defined as follows:

\begin{itemize}[leftmargin=*] \setlength{\itemsep}{-\itemsep}
    \item \textbf{num\_epoch}: total training epochs.
    \item \textbf{batch\_size}: training batch size.
    \item \textbf{optimizer}: the optimizer used for training.
    \item \textbf{adam\_lr}: the learning rate of the optimizer.
    \item \textbf{latent\_dim\_mf}: the embedding dimension for the generalized matrix factorization method.
    \item \textbf{latent\_dim\_mlp}: the embedding dimension for the multi-layer perceptron method.
    \item \textbf{num\_negative}: the number of negative samples for each positive sample.
    \item \textbf{layers}: the dimension parameters of the fully connected layers.
    \item \textbf{l2\_regularization}: the weight coefficient for regularization loss.
    \item \textbf{device\_id}: the computing device used for training.
\end{itemize} 

Note that the basic parameters of ``pmf\_config'', ``vaecf\_config'', and ``ngcf\_config'' are similar to those of ``neumf\_config'', with differences only in the unique configurations specific to each recommendation model. In addition, we provide four folders, i.e., ``original'', ``S-DRO''~\cite{wen2022dro}, ``UFR''~\cite{li2021fairness}, and ``In-Naive'', for comparative and ablation experiments. All configurations are identical to those in our main folder, i.e., ``In-UCDS''.

\subsection{Experimental Environment}

Our source codes are tested in the following environment.

\begin{itemize}[leftmargin=*] \setlength{\itemsep}{-\itemsep}
    \item \textbf{System and Hardware}: Ubuntu 20.04.1 LTS, AMD EPYC 75F3 32-Core Processor (64 threads) @2.95GHz and NVIDIA GeForce RTX 3090.
        \item \textbf{CUDA Toolkit and CuDNN}: Tested with CUDA==11.7 and CuDNN==8.5.0.96.
        \item \textbf{Version and Dependencies}: 
        The main libraries and their versions used in this experiment are listed below:
        
        \texttt{python==3.9.21}

        \texttt{numpy==1.23.0}

        \texttt{torch==1.13.0}

        \texttt{tqdm==4.64.0}

        \texttt{scikit-learn==0.24.2}

        \texttt{pandas==1.5.0}

        \texttt{matplotlib==3.5.1}

        We have packaged them in a file named ``requirements.txt'' in the root directory.
        You can install them by running the command \texttt{pip install -r requirements.txt}.
\end{itemize} 

\section{REPRODUCTION DETAILS}\label{sec:reproducibility}
In this section, we provide detailed instructions on training a recommendation model incorporating the In-UCDS method.
All parameters are pre-configured by default, and no modifications to the code are required.
If you wish to change the recommendation model or adjust other settings, simply modify the corresponding parameters in the command line.
The experiment can be executed using the following command:
\begin{lstlisting}[language=bash]
    cd In-UCDS
    python main.py
\end{lstlisting}

After training is completed, the optimal prediction results will be stored in the ``result'' folder, while the best-performing model weights will be saved in the ``logs'' folder.
To ensure the reliability of the experimental results, we also provide pre-trained model weights, available at:

\begin{flushleft}
    \texttt{\url{https://drive.google.com/drive/folders/17D5sp5mdNOIgRAXohBHU5fBg6-T4BMlk?usp=sharing}} 
\end{flushleft}

In this experiment, the weights of all trained models (including Original and In-UCDS) across four recommendation models and three datasets have been stored in the ``weights'' folder. After downloading the ``weights'' folder, replace the original ``logs'' folder with it, then select the desired recommendation model and dataset for testing, as shown in the following example command:

\begin{lstlisting}[language=bash]
python test.py --model NeuMF --dataset Epinion
\end{lstlisting}

\begin{table*}\centering
\caption{Experimental result. Note that the results of our proposed In-UCDS framework are highlighted in bold, while the best results are marked with an asterisk (*).}
\label{experiment-result-table}
\resizebox{\linewidth}{!}{
\begin{tabular}{ccccccccccccccc}
\hline
\multicolumn{3}{c}{} & \multicolumn{4}{c}{Epinion} & \multicolumn{4}{c}{MovieLens} & \multicolumn{4}{c}{Gowalla}  \\\cmidrule(lr){4-7}\cmidrule(lr){8-11}\cmidrule(lr){12-15}

\multicolumn{3}{c}{} & Overall & Adv. & Disadv. & \UOFMatric  & Overall & Adv. & Disadv. & \UOFMatric  & Overall & Adv. & Disadv. & \UOFMatric  \\ \hline

\multirow{8}{*}{PMF} & \multirow{4}{*}{NDCG} & Original & 0.3783 & 0.4591* & 0.3766 & 0.0128 & 0.4920 & 0.6119 & 0.4857 & 0.1562 & 0.3518 & 0.4480 & 0.3465 & 0.1126 \\

& & S-DRO & 0.3821 & 0.4557 & 0.3806 & 0.0116 & 0.4911 & 0.6114 & 0.4847 & 0.1557 & 0.3546 & 0.4488 & 0.3492 & 0.1090 \\

& & UFR & 0.3765 & 0.4474 & 0.3744 & 0.0083 & 0.4987 & 0.6005 & 0.4867 & 0.1422 & 0.3613 & 0.4325 & 0.3600 & 0.0921\\

& & \M & \textbf{0.3978}* & \textbf{0.4332} & \textbf{0.3823}* & \textbf{0.0041}* & \textbf{0.5050}* & \textbf{0.6302}* & \textbf{0.4984}* & \textbf{0.1325}* & \textbf{0.3826}* & \textbf{0.4546} & \textbf{0.3789}* & \textbf{0.0891}* \\ \cline{2-15}

& \multirow{4}{*}{F1} & Original & 0.1151 & 0.1337* & 0.1140 & 0.0130 & 0.1384 & 0.1569 & 0.1374 & 0.0264 & 0.1050 & 0.1240 & 0.1040 & 0.0225
\\

& & S-DRO & 0.1146 & 0.1311 &0.1137 & 0.0104 & 0.1384 & 0.1569 & 0.1374 & 0.0253 & 0.1058 & 0.1240 & 0.1048 & 0.0192 \\

& & UFR & 0.1120 & 0.1287 & 0.1116 & 0.0094 & 0.1305 & 0.1524 & 0.1300 & 0.0245 & 0.1103 & 0.1205 & 0.1099 & 0.0178 \\

& & \M & \textbf{0.1159*} & \textbf{0.1311} & \textbf{0.1150}* & \textbf{0.0076}* & \textbf{0.1401}* & \textbf{0.1605}* & \textbf{0.1390}* & \textbf{0.0217}* & \textbf{0.1127}* & \textbf{0.1271}* & \textbf{0.1120}* & \textbf{0.0135}*
 \\ \hline

\multirow{8}{*}{VAECF} & \multirow{4}{*}{NDCG} & Original & 0.3035 & 0.3412 & 0.3066 & 0.0357 & 0.5275 & 0.5781 & 0.5233 & 0.0491 & 0.3317 & 0.3917 & 0.3309 & 0.0701 \\

& & S-DRO & 0.3095 & 0.3520 & 0.3050 & 0.0256 & 0.5278 & 0.5709 & 0.5213 & 0.0267 & 0.3405 & 0.4051* & 0.3321 & 0.0359\\

& & UFR & 0.3233 & 0.3376 & 0.3237 & 0.0258 & 0.5132 & 0.5645 & 0.5027 & 0.0208 & 0.3555 & 0.3734 & 0.3510 & 0.0509\\

& & \M & \textbf{0.3567}* & \textbf{0.3705}* & \textbf{0.3579}* & \textbf{0.0158}* & \textbf{0.5616}* & \textbf{0.6018}* & \textbf{0.5594}* & \textbf{0.0191}* & \textbf{0.3931}* & \textbf{0.4034} & \textbf{0.3992}* & \textbf{0.0046}* \\ \cline{2-15}

& \multirow{4}{*}{F1} & Original & 0.0979 & 0.1020 & 0.0947 & 0.0082 & 0.1472 & 0.1532 & 0.1452 & 0.0039 & 0.1013 & 0.1145* & 0.0995 & 0.0108 \\

& & S-DRO & 0.0987 & 0.0993 & 0.0977 & 0.0084 & 0.1465 & 0.1496 & 0.1456 & 0.0049 & 0.1012 & 0.1081 & 0.1010 & 0.0098\\

& & UFR & 0.1002 & 0.1009 & 0.0984 & 0.0077 & 0.1425 & 0.1503 & 0.1427 & 0.0028 & 0.1057 & 0.1120 & 0.1045 & 0.0097\\

& & \M & \textbf{0.1093}* & \textbf{0.1127}*  & \textbf{0.1094}* & \textbf{0.0070}* & \textbf{0.1514}* & \textbf{0.1569}* & \textbf{0.1525}* & \textbf{0.0007}* & \textbf{0.1140}* & \textbf{0.1113} & \textbf{0.1155}* & \textbf{0.0044}* \\ \hline

\multirow{8}{*}{NeuMF} & \multirow{4}{*}{NDCG} & Original & 0.3671 & 0.4078 & 0.3654 & 0.0126 & 0.5407 & 0.6112* & 0.5365 & 0.1095 & 0.3423 & 0.4516 & 0.3364 & 0.1128 \\ 

& & S-DRO & 0.3641 & 0.4074 & 0.3622 & 0.0104* & 0.5337 & 0.6035 & 0.5301 & 0.1053 & 0.3422 & 0.4563* & 0.3366 & 0.1105\\

& & UFR & 0.3741 & 0.4045 & 0.3733 & 0.0117 & 0.5404 & 0.5800 & 0.5428 & 0.0754 & 0.3407 & 0.4379 & 0.3398 & 0.0845\\

& & \M & \textbf{0.4147}* & \textbf{0.4213}* & \textbf{0.4155}* & \textbf{0.0179} & \textbf{0.5740}* & \textbf{0.5455} & \textbf{0.5760}* & \textbf{0.0160}* & \textbf{0.4138}* & \textbf{0.4139} & \textbf{0.4136}* & \textbf{0.0611}*
 \\ \cline{2-15}

& \multirow{4}{*}{F1} & Original & 0.1091 & 0.1140 & 0.1089 & 0.0056 & 0.1483 & 0.1496 & 0.1481 & 0.0150 & 0.1018 & 0.1176 & 0.1010 & 0.0207\\

& & S-DRO & 0.1087 & 0.1167 & 0.1083 & 0.0038 & 0.1489 & 0.1496 & 0.1485 & 0.0108 & 0.1020 & 0.1208* & 0.1013 & 0.0176\\

& & UFR & 0.1102 & 0.1128 & 0.1100 & 0.0037 & 0.1445 & 0.1477 & 0.1443 & 0.0045 & 0.1054 & 0.1127 & 0.1045 & 0.0160\\

& & \M & \textbf{0.1204}* & \textbf{0.1232}* & \textbf{0.1203}* & \textbf{0.0008}* & \textbf{0.1532}* & \textbf{0.1532}* & \textbf{0.1532}* & \textbf{0.0032}* & \textbf{0.1179}* & \textbf{0.1145} & \textbf{0.1179}* & \textbf{0.0028}* \\ \hline

\multirow{8}{*}{NGCF} & \multirow{4}{*}{NDCG} & Original & 0.3691 & 0.3900 & 0.3672 & 0.0135 & 0.4415 & 0.6023* & 0.4462 & 0.1289 & 0.3317 & 0.4389 & 0.3265 & 0.1184 \\

& & S-DRO & 0.3677 & 0.3738 & 0.3707 & 0.0113 & 0.4512 & 0.5977 & 0.4397 & 0.1508 & 0.3410 & 0.4448* & 0.3356 & 0.1095\\

& & UFR & 0.3721 & 0.3865 & 0.3713 & 0.0090 & 0.4512 & 0.5883 & 0.4476 & 0.1105 & 0.3256 & 0.4163 & 0.3199 & 0.1020\\

& & \M & \textbf{0.3979}* & \textbf{0.4061}* & \textbf{0.3974}* & \textbf{0.0073}* & \textbf{0.4739}* & \textbf{0.5955} & \textbf{0.4653}* & \textbf{0.0897}* & \textbf{0.3418}* & \textbf{0.4238} & \textbf{0.3358}* & \textbf{0.0929}* \\ \cline{2-15}

& \multirow{4}{*}{F1} & Original & 0.1085 & 0.1206 & 0.1078 & 0.0106 & 0.1224 & 0.1496 & 0.1216 & 0.0313 & 0.0991 & 0.1176 & 0.1008 & 0.0222 \\

& & S-DRO & 0.1093 & 0.1193 & 0.1086 & 0.0074 & 0.1247 & 0.1532* & 0.1238 & 0.0320 & 0.1015 & 0.1208* & 0.1006 & 0.0237\\

& & UFR & 0.1041 & 0.1179 & 0.1032 & 0.0075 & 0.1187 & 0.1488 & 0.1179 & 0.0218 & 0.0926 & 0.1134 & 0.0978 & 0.0183\\

& & \M & \textbf{0.1149}* & \textbf{0.1272}*  & \textbf{0.1145}* & \textbf{0.0005}* & \textbf{0.1373}* & \textbf{0.1532}* & \textbf{0.1369}* & \textbf{0.0143}* & \textbf{0.1026}* & \textbf{0.1176} & \textbf{0.1018}* & \textbf{0.0172}* \\ \hline

\end{tabular}
}
\end{table*}

\section{EVALUATION EXPERIMENTS}\label{sec:exp}
%
\subsection{Experimental Setup}

\subsubsection{Recommendation Models}
We adopt NeuMF~\cite{he2017neural}, VAECF~\cite{liang2018var}, PMF~\cite{sala2007prob}, and NGCF~\cite{wang2019neural} as the backbone recommendation models.

\subsubsection{Comparison Methods}
We compare our proposed approach with several competing methods, including Original, S-DRO~\cite{wen2022dro}, UFR~\cite{li2021fairness}, and In-Naive.

\subsubsection{Evaluation Metrics}
We employ Normalized Discounted Cumulative Gain (NDCG) and F1-score (F1) to assess the performance of each model, where higher values indicate better recommendation performance. 
The ranking cut-off is set to topK=10. 
Additionally, we utilize \UOFMatric~to measure the recommendation performance gap between advantaged and disadvantaged users, where a smaller value indicates a fairer recommendation performance.

\subsection{Evaluation Results}

\subsubsection{Overall Comparison}
Table~\ref{experiment-result-table} presents the recommendation performance of our proposed method and other comparison methods across the overall, advantaged, and disadvantaged user groups. 

\subsubsection{Ablation Study}
In this ablation study, we select NeuMF as the backbone recommendation model and introduce a simplified variant, named In-Naive, for comparative analysis. 
This method pairs each disadvantaged user with top-ranked advantaged users based on the number of interactions with the same items.
Table~\ref{ablation-study-table} presents the results of our ablation study.

\subsubsection{Effectiveness}
We compare the variation of $L_{fairness}$ in the Original and In-UCDS models as the number of training epochs increases. The results demonstrate that our method effectively narrows the learning gap between advantaged and disadvantaged users. Due to space constraints, we present our comparative results in ``figure.ipynb''.

\begin{table}\centering
  \caption{Ablation study.}
  \label{ablation-study-table}
  \resizebox{\linewidth}{!}{
    \begin{tabular}{cccccccc}
    \hline
& & \multicolumn{2}{c}{Epinion} & \multicolumn{2}{c}{MovieLens} & \multicolumn{2}{c}{Gowalla}  \\\cmidrule(lr){3-4}\cmidrule(lr){5-6}\cmidrule(lr){7-8}

& & Overall & \UOFMatric  & Overall & \UOFMatric  & Overall & \UOFMatric  \\
\hline

\multirow{2}{*}{NDCG} & In-Naive & 0.4031& 0.0204& \textbf{0.5876}& 0.0387& 0.3660& 0.0844\\

& \M & \textbf{0.4147}& \textbf{0.0179}& 0.5740& \textbf{0.0160}& \textbf{0.4138}& \textbf{0.0611}\\ \hline

\multirow{2}{*}{F1} & In-Naive & 0.1176& 0.0016& \textbf{0.1569}& 0.0033& 0.1058& 0.0036\\

& \M & \textbf{0.1204}& \textbf{0.0008}& 0.1532& \textbf{0.0032}& \textbf{0.1179}& \textbf{0.0028}\\

\hline 
    
    \end{tabular}
    }
\end{table}

\section{CONCLUSION}\label{sec:conclusion}
In this paper, we provide the details of the artifacts of the paper ``In-processing User Constrained Dominant Sets for User-Oriented Fairness in Recommender Systems'' for replication.
The artifacts include the datasets and source code used in our experiments.
Based on the experimental setup described above, our proposed framework can be successfully reproduced.


\bibliographystyle{ACM-Reference-Format}
\balance
\bibliography{reference}

\appendix

\end{document}